\documentclass[aps,preprintnumbers,amsmath,amssymb,twocolumn,showpacs]{revtex4}

\usepackage{graphicx}

\newcommand{\fract}[2]{{\textstyle\frac{#1}{#2}}}

\newcommand{\Dslash}{D\hspace*{-7pt}\slash\hspace{2pt}}
\newcommand{\partslash}{\partial\hspace*{-5.5pt}\slash\hspace{2pt}}
\begin{document}

\title{Parity Doubling and  $\mathbf{SU(2)_{L}\times SU(2)_{R}}$ Restoration  in the Hadron Spectrum}

\author{R. L. Jaffe,
Dan Pirjol and Antonello Scardicchio}
\affiliation{ Center for Theoretical Physics, Laboratory for Nuclear Science, and  Department of Physics\\
Massachusetts Institute of Technology, Cambridge, MA 02139, USA}


\begin{abstract}
We construct the most general nonlinear representation of chiral
$SU(2)_L \times SU(2)_R$ broken down spontaneously to the isospin
SU(2), on a pair of hadrons of same spin and isospin and opposite
parity. We show that any such representation is equivalent,
through a hadron field transformation, to two irreducible
representations on two hadrons of opposite parity with different
masses and axial couplings. This implies that chiral symmetry
realized in the Nambu-Goldstone mode does not predict the
existence of degenerate multiplets of hadrons of opposite parity
nor any relations between their couplings or masses.
\end{abstract}
 \vspace*{-\bigskipamount}
\preprint{MIT-CTP-3700}

\maketitle

For massless up and down quarks, QCD has an exact $SU(2)_L \times
SU(2)_R$ chiral symmetry realized in the Nambu-Goldstone mode. The
axial $SU(2)_{A}$ is not a symmetry of the vacuum, and is instead
manifested by the appearance of massless Goldstone bosons, the
pions.  The unbroken isospin subgroup, $SU(2)_{V}$, is realized
explicitly in the Wigner-Weyl mode, by degenerate isospin
multiplets in the mass spectrum.  Small chiral symmetry violating
corrections due to the $u$ and $d$ quark masses can be calculated
in chiral perturbation theory.  An analogous situation holds for
three massless quarks ($u$, $d$, and $s$), albeit with larger
symmetry breaking corrections.  In this paper we treat the two
flavor case, and, for clarity, assume that the $u$ and $d$ quarks
(and the pions) are exactly massless.

The implications of chiral symmetry can be obtained in a
systematic way using effective chiral Lagrangians organized in a
derivative expansion \cite{Cheff,chsymm,Weinberg:1968de,book}.  The
operators in the chiral Lagrangian are the most general chiral
invariants at each order in power counting, constructed from pion
and matter (hadron) fields, which transform nonlinearly under
chiral transformations.  

It has been suggested in the literature that, although chiral
symmetry is broken spontaneously, it can be ``restored'' in
certain sectors of the theory, specifically among highly excited
baryons and mesons.  Because the symmetry includes pseudoscalar
(``axial'') charges, restoration would imply that hadrons of
opposite parity form (approximately) degenerate multiplets.  This
suggestion has been offered as an explanation of ``parity
doubling'', the tendency for hadrons of the same spin and isospin
and opposite parity to have similar masses (for recent
incarnations of these ideas, see Refs.~\cite{paritydoubling}).

Similar ideas have been applied to the quartet of heavy mesons 
constructed by combining a  charm quark with a light quark
system of quantum numbers $s_\ell^{p_{\ell}} = \frac12^+$ and
$\frac12^-$, respectively.  In these pictures chiral
symmetry is not realized explicitly in the spectrum, but states
appear in pairs of opposite parity and their axial couplings are 
predicted to be related \cite{chiraldoubling,chiraldoubling2}.

Explanations of parity   doubling  as manifestations of
chiral symmetry restoration assume, either explicitly or
implicitly, the existence of representations of $SU(2)_{L}\times
SU(2)_{R}$ which include states of opposite parity, and are thus
larger than the irreducible representations of the unbroken group
$SU(2)_{V}$. In this paper we study the most general
nonlinear realizations of chiral symmetry on a pair of hadron
states with opposite parity. The questions we address are:

\begin{itemize}
\item Do nontrivial representations of chiral symmetry exist,
which encompass hadron states of opposite parity? 
\item Assuming
that chiral symmetry {\em is} realized linearly  in a sector of
the hadron spectrum, does the symmetry  imply nontrivial relations among
hadron properties, such as masses or couplings?
\end{itemize}

The answer to both these questions is in the negative. 
Once chiral symmetry is broken spontaneously with the appearance of 
massless pions, it  does not require states of opposite parity to be 
related in any way.  It is possible to write down representations 
involving states of opposite parity that transform into one another 
linearly under chiral transformations.  However chirally invariant 
operators in the effective Lagrangian destroy any relations between 
their masses and coupling constants.  These operators may be suppressed 
for other, dynamical reasons, restoring the relations between coupling 
constants and masses.  However the relations are not a consequence of 
chiral symmetry, but rather a consequence of the dynamical assumptions 
that suppressed the offending operators.

Our results agree with the classic analysis of Coleman {\it et al.}~\cite{CCWZ} of
 nonlinear representations of a Lie
algebra. However,  we believe that a very explicit solution is
instructive, and throws light on the physics of the problem, which
is somewhat obscured in the formal treatment of \cite{CCWZ} 
and by some claims made in the literature\cite{paritydoubling}. 
We will follow closely the treatment and notations of Weinberg~\cite{Weinberg:1968de,book}.

We conclude that parity doubling cannot be a consequence of 
the $SU(2)\times SU(2)$ symmetry of QCD alone.  If it occurs, 
it must be a manifestation of additional dynamics beyond chiral symmetry.     
In a companion paper \cite{JPS} we take a close look at
the data on parity doubling among the baryons.  We find that the
data do confirm a pattern of parity doubling among excited
non-strange baryons (the evidence is much weaker among strange and
doubly strange baryons, where the data are poorer).  Also in 
Ref.~\cite{JPS} we examine the possibility that restoration of the 
$U(1)_{A}$ symmetry of QCD (broken explicitly by the anomaly) is 
the dynamical origin of parity doubling.

We consider the most general realization of chiral symmetry on a
pair of hadrons $B, B'$ with the same isospin $I$ and unspecified
spin. For definiteness, we take $B(B')$ to be parity-even(-odd). A
given realization is identified uniquely through the action of the
generators on the hadron states. The action of the isospin
operators on the hadron states is given by an isospin rotation
$[T^a, B_i] = - t^a_{ij} B_j$, $[T^a, B'_i] = - t^a_{ij}B'_{j}$,
where $t^a$ are $2I+1$ dimensional matrices giving a
representation of SU(2), and $i,j$ are indices labelling the
isospin state of the hadron.

The action of the \emph{axial} charges $X^a$ on the hadron states
is more complicated. When acting on the pion field,
the effect of an axial rotation is written
as~\cite{Weinberg:1968de},
\begin{eqnarray}\label{piaxial}
[X^a, \pi^b] = - if^{{ab}}(\pi) = -i(\delta^{ab} f(\pi^2) + \pi^a
\pi^b g(\pi^2))
\label{pion}
\end{eqnarray}
where $\pi^{2}=\pi^{a}\pi^{a}$ and we have chosen units such that
$f_{\pi}=1$, making our $\pi^a$ equivalent to $\pi^a/f_{\pi}$ in
conventional units. 
Fixing the functions $f,g$ defines a particular choice for the pion field.
Weinberg\cite{Weinberg:1968de} chooses $f(x) = \frac12 (1- x)$ and
$g(x)= 1$, which we also find convenient.
Our  argument does not depend on a particular convention and will
be formulated by keeping $f,g$ completely general.

We turn next to hadron fields other than pions.  We begin with
the most general form of an axial rotation on $B$ and $B'$ that
conserves parity.  Without loss of generality it can be taken to
act homogeneously on their sums and differences, $S  =
B  + B'$ and $D  = B  - B'$.
The fields $S$ and $D$ do not have definite parity, but are
instead transformed into each other, ${P} S  {P}^\dagger = D $.
The action of an axial rotation on these states has the most
general form
\begin{eqnarray}\label{axialgen}
[X^a, N_i] = \{\pm h_1  \delta^{ab} \pm h_2  \pi^a \pi^b + v
\varepsilon_{abc} \pi^c \} t^b_{ij} N_j\,
\end{eqnarray}
where $N=S,D$, the $\pm$ refers to $S$ and $D$ respectively,
and $(h_{1}, h_{2}, v)$ are functions of $\pi^2$  that can be
different for different hadrons. We neglected here a possible term
of the form $w(\pi^2) \pi^a N_i$, which can be eliminated by a
particularly simple field redefinition, $N\to \Phi(\pi^2) N$.

It is clear at this point that we are allowing representations
of the chiral algebra where parity is embedded in a non-trivial
way, {\it i.e.\/} {for which axial rotations} take a hadron $B$ 
into a {\em different} hadron, $B'$, of opposite parity, in addition
to creating pions.
Particular cases of transformations include:
\begin{itemize}
\item $(h_1 ,h_2 ,v ) = (1,0,0)$. This is a linear representation,
  the $(I_{L},0)$ and $(0,I_{R})$ representations for $S$ and $D$ respectively.
  Expressed in terms of the fields with definite parity,
\begin{eqnarray}
\label{lin}
[X^a , B_i] = t^a_{ij} B'_j\,, \qquad [X^a , B'_i] = t^a_{ij} B_j.
\end{eqnarray}
\item  $(h_1,h_2,v) = (0,0,v_0(x))$.  This is the standard
nonlinear (SNL), parity-conserving realization of chiral symmetry on 
the two states $B$ and $B'$ separately,
\begin{eqnarray}
\label{W}
[X^a , N_i] = v_0(\pi^2) \varepsilon_{abc} \pi^c t^b_{ij} N_j\,,
\qquad N = B, B'
\end{eqnarray}
The form of the $v(x)$ function in this case is fixed uniquely by
chiral symmetry \cite{Weinberg:1968de} as $v(x) = v_0(x) \equiv
\frac{1}{x} (\sqrt{f^2(x) +x} - f(x) )$. With our definition of
the pion field, this is $v_0(x) = 1$.
\end{itemize}

We first construct the most general form
of the chiral symmetry realization $(h_{1}, h_{2}, v)$ compatible
with the Lie algebra of the chiral group. We then show that,
through appropriate field transformations (which possibly mix the
hadrons of opposite parity $B, B'$ {and pions}), all such realizations 
are physically equivalent to two decoupled nonlinear
realizations on two isospin multiplets of opposite parity $\tilde
B, \tilde B'$. This proves the absence of nontrivial realizations,
linear or nonlinear, of the chiral symmetry which mix matter
states of opposite parity.

The most general form for the transformation of the hadron fields
eq.~(\ref{axialgen}) under an axial rotation is specified by the
set of three functions $(h_1(x), h_2(x), v(x))$ subject to the
consistency conditions
\begin{eqnarray}
&& 2 v f + h_1^2 + x v^2 = 1\nonumber \\
&&  v g + 2 v' (f + x g) + h_1 h_2 = v^2
\label{jacobi}
\end{eqnarray}
where the $x$ dependence of $f$, $g$, $h_{1}$, $h_{2}$, and $v$
has been suppressed. These equations follow from the Jacobi
identity $[X^a, [X^b, B]] - [X^b, [X^a, B]]  = i\epsilon_{abc}
[T^c, B]$.  The first of eqs.~(\ref{jacobi}) requires that at
fixed $x$, $h_{1}$ and $v$ lie on an ellipse for arbitrary $f$ and
$g$. We write the general solutions of these equations in
terms of an arbitrary function $\theta(x)$, chosen such that
$\theta=0$ corresponds to the SNL solution, eq.~(\ref{W}),
\begin{eqnarray}
\label{soltheta}
&& h_1(x)  = \sqrt{1+f^{2}(x)/x}\ \sin\theta(x)\nonumber  \\
&& v(x)  =\fract1x   \sqrt{f^2(x) +x}  \ \cos\theta(x)  -  f(x)/x
\end{eqnarray}
and $h_{2}(x)$ is determined from the second of eqs.~(\ref{jacobi}). 

\begin{figure}[ht]
\begin{center}
\scalebox{0.62}{\includegraphics{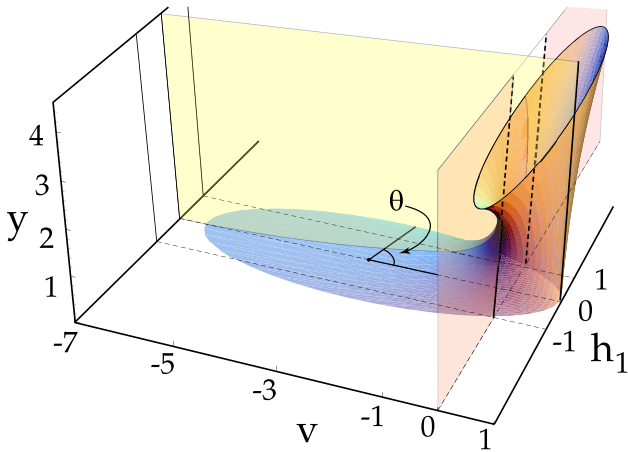}}
\caption{\label{boot} Solutions space for representations of
chiral symmetry that appear to be non-trivial in parity.  $h_{1}$
and $v$ are plotted as functions of $y=2\ln(1+x)$, with the origin 
of the $y$ axis at $y_0>0, y_0 \ll 1$.  At each $y$,
$h_{1}$ and $v$ lie on an ellipse. Any curve $(h_1(y),v(y))$
on this surface corresponds to one particular realization of the chiral
symmetry on a pair of states of opposite parity.
The vertical lines $h_{1}=\pm
1, v=0$ and $h_{1}=0, v=1$ lie on the surface. They are the linear and SNL
representations discussed in the text.  }
\end{center}
\end{figure}

The space spanned by the solutions to eqs.~(\ref{soltheta}) is
shown in Fig.~\ref{boot}  for the particular pion representation 
used here. 
Notice that the surface in Fig.~\ref{boot} is simply connected,
extends all the way from $x=0$ to infinity and it is diffeomorphic
to a cylinder. Any curve $(v(x), h_{1}(x))$ lying on the surface
of Fig.~\ref{boot} and having a unique intersection with each of
the constant-$x$ ellipses, is an allowed nonlinear representation
of $SU(2)_{L}\times SU(2)_{R}$, which mixes the baryons of opposite
parity.
The linear realizations, eqs.~(\ref{lin}) correspond to the vertical 
lines $h_{1}=\pm 1$, $v=0$, or $\theta(x)=\pm \arctan(x/f(x))$, and 
the SNL realization is the vertical line 
$h_{1}=0$, $v=1$, or $\theta(x)=0$.

In the SNL the masses and coupling constants of different multiplets 
of definite isospin and parity are unrelated.  So the questions posed at 
the outset reduce to:  Is there a
redefinition of the  fields $B, B'$ that maps \emph{any} curve
$(v(x), h_{1}(x))$ on the surface of Fig.~(\ref{boot}) to
the SNL realization $\theta(x)=0$?  The redefinition we seek is given by
a transformation of the form
\begin{equation} 
\label{transf}
\tilde S_i =  \Big[ \exp \Big( 2 i\pi^a t^a\Theta_{S}(x)/\sqrt{x}\,
\Big)\Big]_{{ij}} S_j
\end{equation}
and analogously for $D$ (with $\Theta_{S}=\Theta^{*}_{S}$ and $\Theta_{D}=-\Theta_{S}$).   
It is easy to show that the new fields $\tilde
S,\tilde D$ are still a representation of the group with their own
functions $(\tilde h_{1}, \tilde h_{2}, \tilde v)$ satisfying
eq.~(\ref{jacobi}). Hence this solution
can be parameterized by a single angle $\tilde\theta$, which
is furthermore given by $\tilde\theta=\Theta+\theta$. If we choose  
$\Theta(x)=-\theta(x)$, then we have $\tilde\theta(x)=0$. 
Thus
\emph{any solution $(h_{1}(x),h_{2}(x),v(x))$ can be mapped to the
parity conserving standard nonlinear realization of eq.~(\ref{W}), 
corresponding
to $\theta(x)=0$}. Because the surface in Fig.~(\ref{boot}) is
simply connected the function $\Theta$ is continuous and the field
redefinition is always allowed.

We illustrate this formal result by quoting the explicit form of the
field redefinition which takes the baryon fields transforming in the
linear representation eq.~(\ref{lin}) into two decoupled nonlinear
transforming fields $\tilde B, \tilde B'$ with opposite parity
\begin{eqnarray}\label{lin2nl}
\tilde B = \frac{B - 2  i \pi^a t^a B'}
   {\sqrt{1+ \pi^2}} \,, \qquad
\tilde B' = \frac{B' - 2 i \pi^a t^a B}
   {\sqrt{1+  \pi^2}} 
\end{eqnarray} 

Finally we examine the way that the typical physical consequences
of a linear representation of chiral symmetry --- relations among
masses and coupling strengths --- are undone by the presence of
massless pions, and what dynamical assumptions are needed to restore 
them.   We consider the simplest case --- two hadron fields, $B$ and $B'$ 
postulated to obey the linear transformation law, eq.~(\ref{lin}), 
such
that $B$ and $B'$ lie in the  $(I,0) \pm (0,I)$
representations of $SU(2)_L \times SU(2)_R$, respectively.
Under axial
rotations they transform into one another, independently of the
pion field, as if the symmetry were realized in the Wigner-Weyl mode.
Of course, the pions transform non-linearly, by eq.~(\ref{piaxial}).    
The most general effective Lagrangian
invariant under eqs.~(\ref{piaxial}) and (\ref{lin}),
containing only operators of dimension $d \leq 4$, is
\begin{eqnarray}
\label{Llin}
 {\cal L}&=& \bar B i\partslash B + \bar B' i\partslash B' - m_0
(\bar B B + \bar B' B') \\
 &+& m_1  \Big( \bar B\frac{1- \pi^2}{1+\pi^2}B
 - \bar B \frac{4i\pi^a t^a}{1+\pi^2} B'- (B\leftrightarrow B') \Big)\nonumber
\end{eqnarray}
Without loss of generality we assume here and in the following that
$B,B'$ are spin-1/2 baryons.  If the second line of eq.~(\ref{Llin}) 
were ignored, the hadrons described by $B$ and
$B'$ would be degenerate. 
However the term proportional to $m_{1}$, allowed by the nonlinear 
transformations of eqs.~(\ref{piaxial}) and (\ref{lin}), breaks the 
degeneracy of $B$ and $B'$.  The actual physical content of 
${\cal L}$ can more easily be seen by going over to the nonlinearly 
transforming fields $\tilde B, \tilde B'$ defined in eq.~(\ref{lin2nl}).
Expressed in terms of these fields, the
Lagrangian eq.~(\ref{Llin}) assumes the form 
\begin{eqnarray}\label{L2}
{\cal L} &=& \bar {\tilde B} (i\partslash -\varepsilon^{abc}
\pi^a \Dslash \pi^b t^c) \tilde B -
\bar {\tilde B} (\Dslash \pi^a) t^a
{\tilde B}' \nonumber \\
& & \hspace{-1cm} +(\tilde B \leftrightarrow \tilde B') - (m_0 -
m_1) \bar {\tilde B} \tilde B - (m_0 + m_1) \bar {\tilde B}'
\tilde B'
\end{eqnarray}
where $D_\mu \pi^a = 2\partial_\mu \pi^a/(1+\pi^2)$ is the covariant
derivative of the pion field. Note that $\tilde B$ and $\tilde B'$, the 
mass eigenstates, are not degenerate.
 
Invariance of the Lagrangian eq.~(\ref{L2}) under axial
transformations leads  also to predictions concerning the axial charges 
of $B$ and $B'$.  The conserved axial Noether current is
\begin{eqnarray}
\label{Amua}
A_\mu^a = \bar{\tilde  B} \gamma_\mu t^a {\tilde B}' + 
\bar{\tilde B}' \gamma_\mu t^a \tilde B + \mbox{(pion terms)}
\end{eqnarray}
so the axial charges of $B$ and $B'$ vanish, and the off-diagonal $BB'$ 
axial charge is unity. 
These predictions are not disturbed by the term proportional to $m_{1}$.
However, they are invalidated by
further chirally invariant terms involving the covariant derivative of the pion 
field  that
can be added into the Lagrangian.  These can be written in terms of $B$ and $B'$ and 
are invariant when they transform linearly (see eq.~(\ref{lin})) and the pion transforms 
according to eq.~(\ref{piaxial}).  There are three possible terms invariant
under parity, and linear in $D_{\mu}\pi$.  It is easiest to write them in terms 
of the redefined fields,  $\tilde B$ and $\tilde B'$,
\begin{eqnarray}
\delta {\cal L}_{2}&=&c_{2}[  \bar{\tilde B} (\Dslash \pi^a)\gamma_5 t^a \tilde B 
              + \bar{\tilde B}' (\Dslash \pi^a)\gamma_5 t^a \tilde B']\nonumber\\
&+&c_{3}[ \bar{\tilde B} (\Dslash \pi^a) t^a \tilde B'
              + \bar{\tilde B}' (\Dslash \pi^a) t^a \tilde B]\nonumber \\
&+&c_{4}[\bar{\tilde B} (\Dslash \pi^a)\gamma_5 t^a 
\tilde B -
\bar{\tilde B}' (\Dslash \pi^a)\gamma_5 t^a \tilde B']
\end{eqnarray}
The resulting Noether axial current becomes
\begin{eqnarray}\label{Noether}
A_\mu^a &=& (c_2 + c_4) \bar {\tilde B} \gamma_\mu \gamma_5 t^a
\tilde B +
(c_2 - c_4) \bar {\tilde B}' \gamma_\mu \gamma_5 t^a {\tilde B}' \nonumber \\
&+& (1-  c_3) (\bar {\tilde B} \gamma_\mu t^a \tilde B' +
\mbox{h.c.}) + \mbox{(pion terms)}
\end{eqnarray}
and can accomodate any values of the axial matrix elements between the
states $\tilde B, \tilde B'$. At each order in the derivative expansion new 
operators appear, allowed by chiral symmetry, that contribute to the 
$\tilde B-\tilde B'$ mass splitting and change their couplings.

Finally we comment on the generalized Goldberger-Treiman (GT) relations 
that relate the axial charges, pion couplings, and mass differences of 
states with opposite parity.  Regardless of whether they are degenerate 
or are related by chiral transformations, two baryons of opposite parity 
($\tilde B$ and $\tilde B'$) always obey a GT relation
\begin{eqnarray}\label{GT}
g^{+-}_{\pi}=G^{+-}_{A}(M_{\tilde B'}-M_{\tilde B})/f_{\pi}
\end{eqnarray}
where $g^{+-}_{\pi}$ and $G^{+-}_{A}$ are the pion coupling (defined by
${\cal L}_{BB'\pi} = ig^{+-}_{\pi}
\bar{\tilde B}'\pi^a t^a \tilde B + \mbox{h.c.}$)
and transition axial charge for $\tilde B$ and $\tilde B'$.  
This follows simply from requiring 
axial current conservation $q^{\mu}\langle \tilde B'|A^{a}_{\mu}|\tilde B\rangle =0$
and has nothing to do with the transformation properties of $\tilde B$ or 
$\tilde B'$ or the ``restoration'' of chiral symmetry.  
Note that the GT relation does require that the $s$-wave coupling ($g^{+-}_{\pi}$) 
vanishes if $\tilde B'$ and $\tilde B$ are degenerate, as discussed, for example, 
in Ref.~\cite{chiraldoubling2}.   In  our approach eq.~(\ref{GT}) is simply
a restatement of the connection between the axial current, eq.~(\ref{Noether}), 
and the parameters of the Lagrangian, with $G^{A}_{+-} = 1-c_3$. 
Eq.~(\ref{GT}) can be used to extract the axial coupling $G^{A}_{+-}$ from the
$\tilde B'\to \tilde B\pi$ data.

Previous work on parity and chiral doubling neglects
one or more of the constants $\{m_1,c_{2-4 }\}$, which can result in 
considerable predictive power.  However these predictions are not consequences 
of chiral symmetry. Instead they are the result of whatever (often unstated) 
dynamical assumptions enabled the authors to ignore the terms that would have 
invalidated the predictions. Note that, if $c_{2-4}$ are
arbitrarily set to zero, the nonrenormalization of the conserved Noether axial
current implies that they will not be induced by loop corrections to all orders
\cite{JPS}.
This is confirmed by the one-loop results obtained in Ref.~\cite{MS} using 
chiral perturbation theory.

This brings us to the main conclusion of our paper: {\em if one
attempts to realize chiral symmetry in a linear way on a subset of
states in a world with spontaneous symmetry breaking and massless
pions, the chiral symmetry in fact gives no relations among the properties of
these states, such as masses and couplings. }
Such predictions, that are typical of a symmetry realized in the Wigner-Weyl 
mode,  would hold only if certain chirally invariant operators are dynamically 
suppressed.

We close by noting that the arguments in this paper do not
preclude the restoration of chiral symmetry at high temperature or
high chemical potential.  In both cases the restoration occurs if
QCD undergoes a phase transition to a Wigner-Weyl phase, where
there are no massless Nambu-Goldstone bosons. As mentioned in the
introduction, there is some evidence for parity doubling, at least
in the spectrum of non-strange baryons. In Ref.~\cite{JPS} we
examine this evidence and speculate on possible explanations
(including those of Refs.~\cite{Dashen} and \cite{mended}), other
than restoration of $SU(2)_{L}\times SU(2)_{R}$, which as we have
shown, cannot occur in the presence of massless pions.

\vspace{0.2cm}
We thank Andy Cohen and Leonid Glozman, for useful discussions, and Bill Bardeen for comments
on an earlier version of this paper. This work was
supported by the U.S. Department of Energy (D.O.E.) under the 
cooperative research agreement DOE-FC02-94ER40818.\\

\begin{center} 
{\bf \sc Note added after publication}
\end{center}

It has been brought to our attention that some statements in this Letter\cite{Jaffe:2005sq}, can be interpreted as denying the possibility that chiral symmetry can be realized linearly on sets of states \emph{that do not couple to pions}.  This is incorrect:  Indeed, it is obvious that hadrons that do not couple to pions can form linear representations of $SU(2)_L\times SU(2)_R$. Formally, in models of such a world, there would be \emph{two} independent $SU(2)_L\times SU(2)_R$ symmetries, one under which the pions transform non-linearly and the other under which the hadrons in question transform linearly.  We had not envisioned such a radical explanation of parity doubling among hadrons in the 1.5 --- 2.5 GeV region.  If this is the case, then to the extent they are parity doubled, the states in question must not interact by pion emission or absorption, a striking prediction that can be tested experimentally.   This remark applies to the work of L.~Glozman and collaborators in Refs.~\cite{glozman}.  We regret any misunderstanding caused by our omission.


\begin{thebibliography}{19}


\bibitem{Cheff}
S.~Weinberg, Physica, {\bf 96A}, 327 (1979); J.~Gasser and
H.~Leutwyler,
  Nucl.\ Phys.\ B {\bf 250}, 465 (1985).

\bibitem{chsymm} 
W.~A.~Bardeen and B.~W.~Lee,
Phys.\ Rev.\  {\bf 177}, 2389 (1969).

\bibitem{Weinberg:1968de}
  S.~Weinberg,
  Phys.\ Rev.\  {\bf 166}, 1568 (1968).

\bibitem{book}
S.~Weinberg, {\em The Quantum Theory of Fields, Vol. 2: Modern
Applications}.


\bibitem{paritydoubling}
 D.~Jido, T.~Hatsuda and T.~Kunihiro,
  Phys.\ Rev.\ Lett.\  {\bf 84}, 3252 (2000)
  [arXiv:hep-ph/9910375];   D.~Jido, M.~Oka and A.~Hosaka,
  Prog.\ Theor.\ Phys.\  {\bf 106}, 873 (2001)
  [arXiv:hep-ph/0110005];
T.~D.~Cohen and L.~Y.~Glozman,
  Phys.\ Rev.\ D {\bf 65}, 016006 (2002)
  [arXiv:hep-ph/0102206],
  Int.\ J.\ Mod.\ Phys.\ A {\bf 17}, 1327 (2002)
  [arXiv:hep-ph/0201242].
%
\bibitem{chiraldoubling}
  M.~A.~Nowak, M.~Rho and I.~Zahed,
  Phys.\ Rev.\ D {\bf 48}, 4370 (1993)
  [arXiv:hep-ph/9209272],
   M.~A.~Nowak, M.~Rho and I.~Zahed, Acta Phys.\ Polon.\ B {\bf 35}, 2377 (2004)  [arXiv:hep-ph/0307102];
 \bibitem{chiraldoubling2}
 
  W.~A.~Bardeen and C.~T.~Hill,
  Phys.\ Rev.\ D {\bf 49}, 409 (1994)
  [arXiv:hep-ph/9304265];  W.~A.~Bardeen, E.~J.~Eichten and C.~T.~Hill,
  Phys.\ Rev.\ D {\bf 68}, 054024 (2003)
  [arXiv:hep-ph/0305049].
%
\bibitem{CCWZ}
S.~R.~Coleman, J.~Wess and B.~Zumino,
Phys.\ Rev.\  {\bf 177}, 2239 (1969);
C.~G.~Callan, S.~R.~Coleman, J.~Wess and B.~Zumino,
  Phys.\ Rev.\  {\bf 177}, 2247 (1969).
%
\bibitem{JPS} R.~L.~Jaffe, D.~Pirjol and A.~Scardicchio, in preparation.
%
\bibitem{MS} 
T.~Mehen and R.~P.~Springer,
  Phys.\ Rev.\ D {\bf 72}, 034006 (2005)
  [arXiv:hep-ph/0503134].
%
\bibitem{Dashen}
R.~Dashen, Phys.\ Rev.\ {\bf 183}, 1245 (1969).
%
\bibitem{mended}
S.~Weinberg,
Phys.\ Rev.\ Lett.\  {\bf 65}, 1177 (1990).
\bibitem{Jaffe:2005sq}
  R.~L.~Jaffe, D.~Pirjol and A.~Scardicchio, this paper,
  Phys.\ Rev.\ Lett.\  {\bf 96}, 121601 (2006).
\bibitem{glozman} 
  L.~Ya.~Glozman, 
  Phys.\ Lett.\ B {\bf 475}, 329 (2000) [arXiv:hep-ph/9908207], B {\bf 587}, 69 (2004) [arXiv:hep-ph/0312354],
  Int.\ J.\ Mod.\ Phys.\ A  {\bf 21} , 475 (2006) [arXiv:hep-ph/0411281];
  T.~D.~Cohen and L.~Ya.~Glozman, 
  Int.\ J.\ Mod.\ Phys.\ A {\bf 17}, 1327 (2002) [arXiv:hep-ph/0201242];
  L.~Ya.~Glozman  and A.~V.~Nefedief, Phys. Rev. D {\bf 73}, 074018 (2006) [arXiv:hep-ph/0603025].
\end{thebibliography}
\end{document}